\documentclass[useAMS,usenatbib]{mn2e}

\usepackage{epsfig}
\usepackage{longtable}
\usepackage{times} 
\newcommand{\sw}{$Swift$}

\def \inte {{\it INTEGRAL\,}}

\def \sax {{\it BeppoSAX}}
\def \sw {{\it Swift}}
\def \src {\mbox{SAX~J1818.6$-$1703}}

\def \hcm {\hbox {\ifmmode $ atom cm$^{-2}\else atom cm$^{-2}$\fi}}

\def \sax{{\it BeppoSAX}}
\def \ATel {Astron.\ Tel.}
\def \apj {ApJ}
\def \apjl {ApJL}
\def \apjs {ApJS}\def \aap {A\&A}

\def \mnras {MNRAS}
\def \iaucirc {IAU Circ.}
\def \aaps {A\&AS}

\title[Broad-band study of SAX~J1818.6$-$1703]{The first broad-band X--ray study of the Supergiant Fast X--ray Transient \src\ in outburst}
\author[L.\ Sidoli et al.]{L.\ Sidoli,$^{1}$ P.\ Romano,$^{2}$ P.\ Esposito,$^{1,3}$  V.\ La Parola,$^{2}$ J.A.~Kennea,$^{4}$ H.A.\ Krimm,$^{5,6}$ 
\newauthor   M.M.~Chester,$^{4}$ A.~Bazzano,$^{7}$  D.N.~Burrows,$^{4}$  N.~Gehrels$^{5}$  \\
$^{1}$INAF, Istituto di Astrofisica Spaziale e Fisica Cosmica,
	Via E.\ Bassini 15,   I-20133 Milano,  Italy\\
$^{2}$INAF, Istituto di Astrofisica Spaziale e Fisica Cosmica,
        Via U.\ La Malfa 153, I-90146 Palermo, Italy\\
$^{3}$INFN, Sezione di Pavia,
        Via A.\ Bassi 6, I-27100, Pavia, Italy \\
$^{4}$Department of Astronomy and Astrophysics, Pennsylvania State 
             University, University Park, PA 16802, USA\\
$^{5}$NASA/Goddard Space Flight Center, Greenbelt, MD 20771, USA\\
$^{6}$Universities Space Research Association, Columbia, MD, USA \\
$^{7}$INAF, Istituto di Astrofisica Spaziale e Fisica Cosmica,
        Via Fosso del Cavaliere 100, I-00133, Roma, Italy \\
}

\begin{document}

\date{}

\pagerange{\pageref{firstpage}--\pageref{lastpage}} \pubyear{2009}

\maketitle

\label{firstpage}

\begin{abstract}
The Supergiant Fast X--ray Transient (SFXT) SAX~J1818.6$-$1703 underwent an outburst 
on 2009 May 6 and was observed with \sw. 
We report on these observations which, for the first time, allow us to study
the broad-band spectrum from soft to hard X--rays  of this source. 
No X--ray spectral information was available on this source before the \sw\ monitoring.
The spectrum can be deconvolved well with models usually adopted 
to describe the emission from High Mass X--ray Binary X--ray pulsars, and
is characterized by a very high absorption, a flat power law 
(photon index $\sim$0.1--0.5) and a cutoff at about 7--12~keV. 
Alternatively, the \src\ emission can be described 
with a Comptonized emission from a cold and optically thick corona, 
with an electron temperature 
$kT_{e}$=5--7~keV, a hot seed photon temperature, $kT_{0}$, of 1.3--1.4~keV,
and an optical depth for the Comptonizing plasma, $\tau$, of about 10. 
The 1--100 keV luminosity at the peak of the flare is 
3$\times$10$^{36}$~erg~s$^{-1}$ (assuming the optical counterpart distance of 2.5~kpc). 
These properties of \src\ resemble those of the prototype of the SFXT class, XTE~J1739--302.
The monitoring with \sw/XRT reveals an outburst duration of about 5~days, 
similarly to other members of the class of SFXTs, 
confirming \src\ as a member of this class.
\end{abstract}



	\section{Introduction\label{sax1818:intro}}

Supergiant Fast X-ray Transients (SFXTs) are transient X--ray sources in binary systems  
composed of a compact object and a blue supergiant companion. Although some of them
were discovered before the \inte\ satellite launch in 2002, this new class of High Mass X--ray Binaries (HMXBs)
was recognized only after several new peculiar X--ray transients had been discovered  
as a result of 
the Galactic plane survey performed by \inte\  (\citealt{Sguera2005}, \citealt{Negueruela2006}).
The members of this new class of sources  (about 10, with $\sim$20 candidates)  display 
apparently short outbursts (as observed with \inte), 
characterized by a few hour duration flares and by a high dynamic range 
(1,\,000--100,\,000) between the quiescence (10$^{32}$~erg~s$^{-1}$) 
and the flare peak (10$^{36}$--10$^{37}$~erg~s$^{-1}$). 
Spectral properties are very similar to those of 
High Mass X--ray Binary pulsars  \citep{Romano2008:sfxts_paperII}.

Their long-term X--ray properties have been investigated thanks to a 
monitoring campaign with \sw\ \citep[still ongoing, see][]{Romano2009:sfxts_paperV}, 
which is unveiling several new properties of a sample of SFXTs.
This \sw\ campaign has demonstrated that the SFXTs short flares are 
part of a longer outburst phase lasting days 
\citep{Sidoli2009:sfxts_paperIII} and they spend most of their lifetime in accretion at an intermediate level
(10$^{34}$~erg~s$^{-1}$), instead of staying in quiescence 
(as was previously thought; \citealt{Sidoli2008:sfxts_paperI}).
The actual mechanism responsible for the outbursts is still unknown 
(see \citealt{Sidoli2009:cospar} for a review of the different proposed scenarios).

\src\ is an X--ray source classified as a Supergiant Fast X-ray Transient 
because of its transient X--ray activity, its high dynamic range \citep{Bozzo2008:atel1493} 
and its association with a blue supergiant star located at 2.5~kpc
(O9--B1 type, \citealt{Negueruela2006:aTel831}, \citealt{Negueruela2007sax1818}), 
thanks to the sub-arcsecond X--ray position obtained using 
$Chandra$ \citep{zand2006:aTel915}.
\src\ was discovered with the Wide Field Cameras on board \sax\  \citep{zand1998}, and showed 
several more bright flares lasting 1--3 hours as observed 
with IBIS/ISGRI on board \inte\  (\citealt{Grebenev2005}, \citealt{Sguera2005}), reaching $\sim$200 mCrab
at the flare peak (18--45 keV), and with ASM on board {\it RXTE} \citep{Sguera2005}. 
Further and more recent activity has been caught both 
by \sw/BAT \citep{Barthelmy2008GCN7419} and by \inte\ \citep{Grebenev2008:atel1482}.
A periodicity (likely orbital) of 30-days was discovered (\citealt{Bird2009}, \citealt{Zurita2009})
from the analysis of available \sw/BAT and \inte\ data, suggesting an eccentric orbit ($e$$\sim$0.3--0.4)
and an  outburst duration of around 4--6~days.

Here we report on the latest outburst from this source, observed 
with \sw\ starting on 2009 May 6 \citep{Romano2009:atel2044}. 
These \sw\ observations allow us to perform the first
broad-band spectroscopy of this SFXT, simultaneously from soft to hard X--rays. 

 	 \section{ Observations and Data Reduction\label{sax1818:dataredu}}

\begin{figure}
\begin{center}
\vspace{-1truecm}
\includegraphics*[angle=0,scale=0.42]{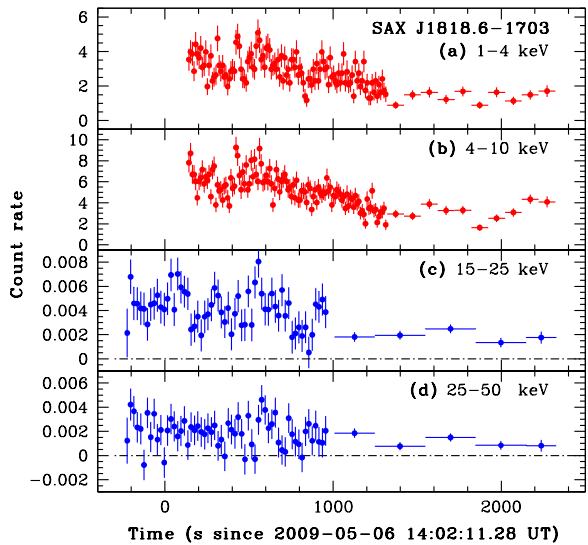}
\end{center}
\vspace{-0.5truecm}
\caption{XRT and BAT light curves of the 2009 May 6 outburst 
in units of count s$^{-1}$ and count s$^{-1}$ detector$^{-1}$, respectively, 
showing data collected through the AT observation.  
The finer sampling points correspond to XRT/WT data (panels a and b) 
and BAT data in event mode (panels c and d); 
the remainder are in XRT/PC data and  survey mode BAT data. 
}
\label{sax1818:fig:lcv_allbands}
\end{figure}

\src\  triggered the \sw/BAT three times since the start of the mission,
on 2007 October 16 at 04:20:16 UT (trigger 294385), 
on 2008 March 15 at 15:54:45 UT \citep[trigger 306379,][]{Barthelmy2008GCN7419}, and 
on 2009 May 6 at 14:03:27 UT \citep[trigger 351323,][]{Romano2009:atel2044}. 
In the first two instances \sw\ did not perform a slew, so no narrow-field instrument (NFI)
observations are available. 

During the flare of 2009 May 6, the subject of this paper, 
\sw\ executed an immediate slew, so that the XRT began observing the
field of \src\ about 132 seconds after the BAT trigger. 
The automated target (AT) observation lasted for one orbit (until approximately 2300 \,s 
after the trigger).  
Follow-up target of opportunity (ToO) observations for 10\,ks were obtained by means 
of a Cycle 5 proposal (PI P.\ Romano; sequences 00031409001--005, see Table~\ref{sax1818:tab:obs}) 
and regular ToO observations (00031409006--010). 
The data cover the first 14\,d since the beginning of the outburst.  

The BAT data were analysed using the standard BAT analysis 
software distributed within {\sc ftools}. 
Mask-tagged BAT light curves were created in the standard 4 energy bands, 
15--25, 25--50, 50--100, 100--150 keV, and rebinned to achieve a signal-to-noise (S/N) of 10
(see Fig.~\ref{sax1818:fig:lcv_allbands}c, d).
BAT mask-weighted spectra were extracted from event files over the time 
interval simultaneous with XRT/WT data (see Table~\ref{sax1818:tab:obs})
and were rebinned to a S/N of 3. 
Response matrices were generated with {\sc batdrmgen}. 
For our spectral fitting ({\sc xspec} v11.3.2) 
we applied an energy-dependent systematic error. 
Furthermore, survey data products, in the form of Detector Plane
Histograms (DPH) with typical integration time of 
$\sim 300$\,s, are available, and were also analysed with the
standard BAT software.

The XRT data were processed with standard procedures 
({\sc xrtpipeline} v0.12.1), filtering and screening criteria by using 
{\sc ftools} in the {\sc Heasoft} package (v.6.6.1).  
We considered both WT and PC data, selected event grades 0--2 and 0--12, respectively 
(\citealt{Burrows2005:XRTmn}). 
We corrected for pile-up when appropriate
by determining the size of the point spread function (PSF) core affected 
by comparing the observed and nominal PSF 
\citep{vaughan2006:050315mn,Romano2006:060124mn},
and excluding from the analysis all the events that fell within that
region.   
To account for the background, we also extracted events within 
source-free regions. 
Exposure maps were generated with the task {\tt xrtexpomap}. 
Ancillary response files were generated with {\sc xrtmkarf},
to account for different extraction regions, vignetting, and
PSF corrections. We used the spectral redistribution matrices v011 in CALDB.

All quoted uncertainties are given at 90\,\% confidence level for 
one interesting parameter unless otherwise stated. 
The spectral indices are parameterized as  
$F_{\nu} \propto \nu^{-\alpha}$, 
where $F_{\nu}$ (erg cm$^{-2}$ s$^{-1}$ Hz$^{-1}$) is the 
flux density as a function of frequency $\nu$; 
we adopt $\Gamma = \alpha +1$ as the photon index, 
$N(E) \propto E^{-\Gamma}$ (ph cm$^{-2}$ s$^{-1}$ keV$^{-1}$). 
Times in the light curves and the text are referred to the trigger time.

\begin{figure}
\begin{center}
\includegraphics*[angle=270,scale=0.35]{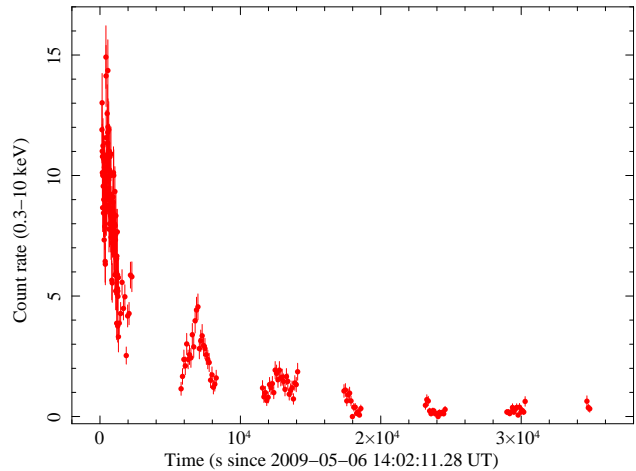}
\end{center}
\caption{\sw/XRT light curve of \src\ in the energy range 0.3--10 keV, on a linear scale, 
during the first day of the bright outburst, showing data collected through 
both AT and GI ToO observations.
}
\label{sax1818:fig:xrtgi}
\end{figure}

 \begin{table*}
 \begin{center}
 \caption{Summary of the {\it Swift} observations.\label{sax1818:tab:obs} }
 \begin{tabular}{llllll}
 \hline
 \noalign{\smallskip}
 Sequence   & Obs/Mode  & Start time  (UT)  & End time   (UT) & Exposure & Time since trigger   \\ 
            &           & (yyyy-mm-dd hh:mm:ss)  & (yyyy-mm-dd hh:mm:ss)  &(s)  & (s)       \\
  \noalign{\smallskip}
 \hline
 \noalign{\smallskip}
00351323000	&	BAT/event &       2009-05-06 13:58:16     &       2009-05-06 14:18:18     &       1202    &       -239     \\ 
00351323000	&	BAT/survey &      2009-05-06 14:19:03     &       2009-05-06 14:57:42     &       1319    &       1,\,008     \\ 
00351323000	&	XRT/WT    &       2009-05-06 14:04:32     &       2009-05-06 14:24:15     &       1183    &       138     \\
00351323000	&	XRT/PC    &       2009-05-06 14:24:17     &       2009-05-06 14:40:57     &       977     &       1,\,322    \\
00031409001	&	XRT/PC    &       2009-05-06 15:37:42     &       2009-05-06 23:44:34     &       9859    &       5,\,728    \\
00031409002	&	XRT/PC    &       2009-05-07 15:45:00     &       2009-05-07 18:02:12     &       4924    &       92,\,565   \\
00031409003	&	XRT/WT    &       2009-05-08 01:41:00     &       2009-05-08 17:44:02     &       168     &       128,\,325  \\
00031409003	&	XRT/PC    &       2009-05-08 01:41:11     &       2009-05-08 17:48:57     &       4782    &       128,\,336  \\
00031409004	&	XRT/PC    &       2009-05-09 03:13:50     &       2009-05-09 17:46:57     &       5919    &       220,\,296  \\
00031409005	&	XRT/PC    &       2009-05-10 06:30:43     &       2009-05-10 09:48:41     &       928     &       318,\,508  \\
00031409006	&	XRT/PC    &	  2009-05-15 21:43:43	  &	  2009-05-15 21:54:58	  &	  665	  &	  805,\,289  \\
00031409007	&	XRT/PC    &	  2009-05-16 13:30:33	  &	  2009-05-16 15:14:57	  &	  1083    &	  862,\,099  \\
00031409008	&	XRT/PC    &	  2009-05-17 20:24:37	  &	  2009-05-17 22:10:58	  &	  1783    &	  973,\,343  \\
00031409009	&	XRT/PC    &	  2009-05-18 15:34:27	  &	  2009-05-18 17:13:57	  &	  1722    &	  1,\,042,\,332 \\
00031409010	&	XRT/PC    &	  2009-05-19 15:34:28	  &	  2009-05-19 17:20:57	  &	  1899    &	  1,\,128,\,734 \\
  \noalign{\smallskip} 
  \hline
  \end{tabular}
  \end{center}
  \end{table*}

 	 \section{Results \label{sax1818:results} }

Fig.~\ref{sax1818:fig:lcv_allbands} shows the light curves 
during the brightest part of the outburst in different energy bands. 
While the BAT count rate was fairly constant throughout the AT observation,
the XRT count rate shows a decreasing trend from a maximum of $\sim 15$ counts s$^{-1}$
(in the 0.2--10\,keV band, see Fig.~\ref{sax1818:fig:xrtgi}) 
with several flares superimposed.  
The X-ray decreasing trend is even more evident in Fig.~\ref{sax1818:fig:xrtgi}, 
which shows the XRT light curve in the 0.3--10\,keV band  on a linear scale 
throughout the first day (AT and GI ToO) of observations. 
The XRT decay time during the first day of the outburst can be fitted 
with an exponential function, with e-folding time, $\tau_{\rm e}$, of 3800$\pm{100}$~s.
Note however that this gives only an idea of the general trend, on top of a lot of
short term variability.

The XRT spectra of the first six temporal sequences reported in Table~\ref{sax1818:tab:obs} have
been analysed separately, to search for variability of the spectral parameters along the 
bright phase of the outburst. The remaining sequences only provided 3$\sigma$ upper limits. 
In Table~\ref{sax1818:tab:xrtspec} we report the spectroscopy results obtained when fitting
the spectra with simple models, an absorbed power law or blackbody. 
More complex deconvolutions of the 1--10~keV spectra are not required by the data.
No spectral variability is evident, within the large uncertainties.

\begin{figure}
\begin{center}
\includegraphics*[angle=270,scale=0.35]{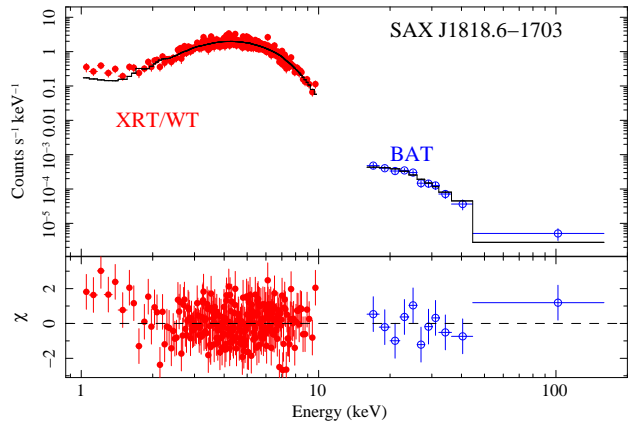}
\vspace{0cm}
\end{center}
\caption{Spectroscopy of the 2009 May 6 outburst. 
		{\bf Top:} simultaneous XRT/WT (filled red circles) and BAT (empty blue circles) data 
			fit with an absorbed Comptonization emission ({\sc bmc} model in {\sc xspec}). 
		{\bf Bottom:} the residuals of the fit (in units of standard deviations). 
}
\label{sax1818:fig:meanspec}
\end{figure}

We extracted simultaneous BAT and XRT spectra, between 138 and 937\,s 
since the BAT trigger and performed a joint fit in the 1--10\,keV and 
14--150\,keV energy bands for XRT and BAT, respectively. 
A constant factor was included to allow for normalization uncertainties between the two
instruments (always resulted within the usual range).
The X--ray spectrum is highly absorbed ($N_{\rm H}$$\sim$5--7$\times$10$^{22}$~cm$^{-2}$) 
and is very well fit with models we already adopted to describe other SFXTs wide-band spectra 
\citep{Sidoli2009:sfxts_paperIV}: power laws with high energy 
cutoffs ({\sc cutoffpl} in {\sc xspec}), 
or Comptonization models [{\sc Comptt}  \citep{Titarchuk1994} or the  
{\sc bmc} model  \citep{TMK1996} in {\sc xspec}].
The results of the broad-band spectroscopy are summarized in Table~\ref{sax1818:tab:broadspec}
(also see Fig.~\ref{sax1818:fig:meanspec}). 

The {\sc bmc} model is composed of a blackbody (BB) and its Comptonization 
and it is not limited 
to the thermal Comptonization case ({\sc Comptt}) but also can account  for 
dynamical (bulk) Comptonization due to the converging flow. 
The spectral parameters of the {\sc bmc} model 
are the black-body (BB) color temperature, $kT_{\rm BB}$, the spectral 
index $\alpha$  (overall Comptonization efficiency 
related to an observable 
quantity in the photon spectrum of the data), 
and the logarithm of the illuminating factor $A$, $\log A$ 
(an indication of the fraction of the up-scattered BB photons with 
respect to the BB seed photons directly visible). 
In \src\ $\log A$ could not be constrained and $\alpha$ was only determined with large uncertainties.
Adopting the {\sc Comptt} model, the properties of the Comptonizing
corona could be constrained well with an an electron temperature $kT_{e}$$\sim$5--7~keV 
and an optical depth $\tau$$\sim$10 (in a spherical geometry).

For the timing analysis we only used the XRT observations in which the source was 
significantly detected. 
After correcting the photon arrival times to the Solar system barycenter, 
we extracted the source events with the same selection criteria used for the spectroscopy.
We searched each single dataset for coherent sub-orbital pulsations down to a minimum 
period corresponding to the Nyquist limit (3.532\,ms and 5.0146\,s for 
the data collected in WT and in PC modes, respectively) using the $Z^2_n$ test \citep{Buccheri1983}, 
with the number of harmonics $n$ being varied from 1 to 4. 
The period search step size was chosen in each dataset in order to oversample 
the Fourier period resolution ($\frac{1}{2}P^2/T_{\rm{obs}}$) by a factor of 5. 
No statistically significant signal was detected. 
The upper limit on the 
pulsed fraction (defined as the semi-amplitude of the sinusoidal modulation 
divided by the mean count rate), computed according to \citet{Vaughan1994}, 
is 12\,\% at the 99\,\% confidence level and for periods between 3.5\,ms and 200\,s.  
In order to search for pulsed signals at long periods ($P>200$ s), 
we analysed together all the data by computing a 
Lomb-Scargle periodogram \citep{Scargle1982}. 
Again, we did not detect any 
significant signal for periods up to $\sim$1000 s,
with an upper limit on the pulsed fraction of about 25\,\% (at the 99\,\% confidence level).

\begin{figure}
\begin{center}
\includegraphics*[angle=270,scale=0.35]{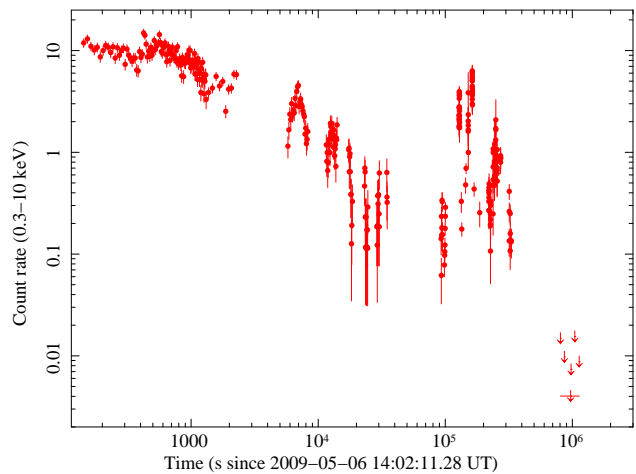}
\vspace{0cm}
\end{center}
\caption{\sw/XRT light curve of \src\ in the energy range 0.3--10 keV, during the whole 
campaign. The four day gap before the 3~$\sigma$ upper limits is due to the source being Moon constrained.
The data shown were collected through both AT and GI ToO, and regular ToO observations.
}
\label{sax1818:fig:full_xrtlcv}
\end{figure}

 	 \section{Discussion \label{sax1818:discussion}}

\src\ is one of the four SFXTs  where a period, 
very likely of orbital origin, has been detected (P$_{\rm orb}$).
Such periods have been so far determined  either through observations of 
periodically recurrent outbursts 
[IGR~J11215--5952, \citet{SidoliPM2006}, \citet{Romano2009:11215_2008}; P$_{\rm orb}$$\sim$165~days]  
or through a timing analysis of the available hard X--ray data [IGR~J18483--0311, \citet{Sguera2007}, P$_{\rm orb}$$\sim$18.5~days; 
IGR~J16479--4514, \citet{Jain2009:16479}, P$_{\rm orb}$$\sim$3.3~days]. 
For a  fifth SFXT, IGR~J08408--4503, we inferred an orbital period of $\sim$35~days from the 
times of the outbursts and the duration of the flares \citep{Romano2009:sfxts_paper08408}. 

The \src\ orbital period of 30~days has been independently obtained  by \citet{Bird2009} and by \citet{Zurita2009}.
Assuming as orbital phase zero the same as Bird et al. (MJD~54540.659), 
the outburst we analyse here (BAT trigger start time) occurred at orbital phase $\phi$=0.90$\pm{0.05}$ (1 $\sigma$ error),
consistent with the folded light curve and the 4--6 days outburst duration reported by both authors.
The new outburst from \src\ we are reporting here indeed 
lasted about 5~days (see Fig.~\ref{sax1818:fig:full_xrtlcv}),
which is also similar to the one
in other SFXTs we have been monitoring 
in the last  year with \sw\ (\citealt{Sidoli2009:sfxts_paperIV}, \citealt{Romano2009:sfxts_paperV}) and
to the outburst duration in IGR~J11215--5952 \citep{Romano2007}.

After the BAT trigger, the whole outburst evolution and the decline phase could be monitored with \sw/XRT.
The source displays a dynamical range of more than 3000 (see Fig.~\ref{sax1818:fig:full_xrtlcv}) 
and a multiple-flaring behaviour previously seen in other SFXTs \citep{Romano2009:sfxts_paper08408}. 
The temporally resolved spectroscopy with XRT did not result in variability of the spectral parameters
in the 1--10~keV range, within the uncertainties, except for the obvious change in the blackbody radius,
when fitting the soft X--rays with a single absorbed blackbody (see Table~\ref{sax1818:tab:xrtspec}).
No variability in the absorbing column density could be detected along the outburst, within the uncertainties.

For the first time we obtained broad-band spectroscopy from soft to hard X--rays of this SFXT,
allowing us to compare it with wide-band spectra from a few other SFXTs. 
Indeed, although \src\ has been observed with \inte\ \citep{Sguera2005} in the past,
only the high energy part was available ($>20$~keV) from the IBIS instrument, 
and the source spectrum has not been obtained (probably because of the low statistics).
The \sw/XRT and BAT joint simultaneous spectrum during the \src\ outburst could be fit very well with a cutoff power law 
({\sc cutoffpl} in {\sc xspec}) or with Comptonized emission models. 
It is highly absorbed, shows a flat power law (photon index in the range 0.1--0.5), 
and displays a cut-off constrained betwen 7 and $\sim12$~keV (Table~\ref{sax1818:tab:broadspec}).
The use of the  {\sc Comptt} model quantifies the physical conditions of the Comptonizing plasma,
resulting in a temperature $kT_{e}$ (5--7~keV) for the Comptonizing plasma and the 
optical depth $\tau$ of the spherical corona ($\tau$$\sim$10). 
The 1--100~keV luminosity is $3\times10^{36}$~erg~s$^{-1}$ (assuming 2.5~kpc, \citealt{Negueruela2007sax1818}).
These \src\ broad-band properties are reminiscent of the X--ray spectral shape
of the prototype of the SFXT class, XTE~J1739--302 (\citealt{Sidoli2009:sfxts_paperIII}, \citealt{Sidoli2009:sfxts_paperIV}).
The two sources indeed display a similarly high absorption, similar X--ray luminosities, energy cutoff values,
and hot seed photon temperatures, $kT_{0}$, of 1.3--1.4~keV. 
These spectral parameters indicate the presence of a cold and optically thick corona, 
which is also similar to what observed in bright 
neutron star low mass X--ray binaries \citep{Paizis2006}.
No statistically significant pulsations have been found in the \sw/XRT data. 

The duration of the outbursts, about 5 days, is also similar to the one
 observed in other SFXTs we have been monitoring 
in the last  year with \sw\ (\citealt{Sidoli2009:sfxts_paperIV}, \citealt{Romano2009:sfxts_paperV}) and
to the outburst duration in IGR~J11215--5952 \citep{Romano2007}.
All these observed properties, together with the broad-band spectrum we  observe now for the first time,
confirm the fact that \src\ belongs to the class of SFXTs.

\section*{Acknowledgments}

We thank the {\it Swift} team duty scientists and science planners. 
We also thank the remainder of the {\it Swift} XRT and BAT teams,
S.\ Barthelmy in particular, for their invaluable help and support. 
This work was supported in Italy by contracts ASI I/088/06/0 and I/023/05/0, at
PSU by NASA contract NAS5-00136.  HAK was supported by the {\it Swift } project.

 \begin{table*} 	
 \begin{center} 	
 \caption{XRT spectroscopy. \label{sax1818:tab:xrtspec}} 	
 \label{} 	
 \begin{tabular}{llllll} 
 \hline 
 \noalign{\smallskip} 
 & 	Power-law&              &           &   \\
Sequence   & $N_{\rm H}$  & $\Gamma$  & Unabsorbed Flux$^a$ & $\chi^{2}_{\rm red}/dof$  & Times \\ 
           &   (10$^{22}$~cm$^{-2}$)           &           & (1--10\,keV)    &  (since trigger) \\
 \hline 
00351323000 WT	&	$7.9^{+0.6}_{-0.6}$	&	$1.1^{+0.1}_{-0.1}$	&   $1.9^{+0.1}_{-0.1} \times10^{-9}$    & 1.0/329& 138--1,\,320 \\
00351323000 PC	&	$9.6^{+2.1}_{-1.8}$	&	$1.1^{+0.4}_{-0.3}$	&   $1.3^{+0.2}_{-0.1} \times10^{-9}$    & 1.0/50& 1,\,322--2,\,299 \\
00031409001 PC	&	$8.1^{+1.0}_{-0.9}$	&	$1.7^{+0.2}_{-0.2}$	&   $3.3^{+0.4}_{-0.3} \times10^{-10}$    & 0.8/112&	5,\,728--34,\,923 \\
00031409002 PC	&	$11.0^{+3.3}_{-2.5}$	&	$1.9^{+0.6}_{-0.5}$	&   $5.4^{+2.9}_{-5.4} \times10^{-11}$    & 0.7/24& 92,\,565--100,\,780 \\
00031409003 WT	&	$7.2^{+1.9}_{-1.6}$	&	$1.9^{+0.5}_{-0.5}$	&   $9.7^{+0.7}_{-2.0} \times10^{-10}$ & 0.9/23&128,\,325--186,\,108   \\
00031409003 PC	&	$8.1_{-1.2}^{+1.4}$	&	$1.5_{-0.3}^{+0.3}$	&   $2.7^{+0.3}_{-0.3} \times10^{-10}$ & 0.7/61&128,\,336--186,\,388    \\ 
00031409004 PC	&	$8.7^{+2.2}_{-1.7}$	&	$1.4^{+0.4}_{-0.3}$	&   $1.5^{+0.3}_{-0.8} \times10^{-10}$ & 0.9/46&220,\,296--272,\,679   \\
00031409005 PC	&	$9.8^{+7.1}_{-1.3}$	&	$0.9^{+1.6}_{-1.3}$	&   $5.7^{+0.8}_{-5.7} \times10^{-11}$ & 0.7/5&318,\,508--330,\,367    \\
 \hline 
 & Blackbody	&              &           & \\
Sequence   & $N_{\rm H}$  & $kT$  & Unabsorbed Flux$^a$ & $\chi^{2}_{\rm red}/dof$ &  $R_{\rm BB}$    \\ 
           &  (10$^{22}$~cm$^{-2}$) &   (keV)        & (1--10\,keV)    &  & (km)$^b$ \\
 \hline 
00351323000 WT	&$4.8^{+0.4}_{-0.4}$&   $2.1^{+0.1}_{-0.1}$ &$1.4 ^{+0.1}_{-0.1} \times10^{-9}$  &    1.0/329	  &$0.8^{+0.1}_{-0.1}$ \\
00351323000 PC	&$6.4^{+1.4}_{-1.2}$&   $2.1^{+0.3}_{-0.3}$ &$9.3 ^{+0.2}_{-0.1} \times10^{-10}$  &    1.0/50	  &$0.6^{+ 0.2}_{- 0.1}$ \\
00031409001 PC	&$4.7^{+0.6}_{-0.5}$&   $1.6^{+0.1}_{-0.1}$ &$2.0 ^{+0.1}_{-0.1} \times10^{-10}$  &    0.8/112	  &$0.4^{+0.1}_{-0.1}$ \\
00031409002 PC	&$6.6^{+2.2}_{-1.7}$&   $1.6^{+0.3}_{-0.2}$ &$2.3 ^{+0.8}_{-0.8}\times10^{-11}$  &    0.7/24	  &$0.2^{+0.1}_{-0.1}$ \\
00031409003 WT	&$4.0^{+1.2}_{-1.0}$&   $1.4^{+0.2}_{-0.2}$ &$5.2 ^{+0.3}_{-0.1}\times10^{-10}$  &    1.0/23	  &$0.9^{+ 0.4}_{- 0.2}$    \\
00031409003 PC	&$4.9_{-0.8}^{+0.9}$&   $1.8_{-0.2}^{+0.2}$ &$1.7 ^{+0.3}_{-0.5} \times10^{-10}$  &    0.9/61          &$0.4^{+0.1}_{-0.2}$    \\ 
00031409004 PC	&$5.2^{+1.3}_{-1.1}$&   $1.8^{+0.2}_{-0.2}$ &$9.7 ^{+0.3}_{-0.3} \times10^{-11}$  &    0.8/46  	  &$0.3 ^{+0.1}_{-0.1}$    \\
00031409005 PC	&$6.7^{+4.8}_{-3.8}$&   $2.2^{+3.7}_{-0.9}$ &$4.2 ^{+2.4}_{-4.2} \times10^{-11}$ &    0.8/5	          & $ <0.3 $    \\

  \noalign{\smallskip}
  \hline
  \end{tabular}
  \end{center}
  \begin{list}{}{} 
  \item[$^{\mathrm{a}}$ Fluxes (corrected for the absorption) are in units of erg~cm$^{-2}$~s$^{-1}$.] 
  \item[$^{\mathrm{b}}$ Blackbody radii are in units of km, assuming the optical counterpart distance of 2.5~kpc.] 
  \end{list} 
  \end{table*} 

 \begin{table*}
 \begin{center}
 \caption{Spectral fits of simultaneous XRT and BAT data of \src.}
 \label{sax1818:tab:broadspec}
 \begin{tabular}{lrrrrrrrr}
 \hline
 \noalign{\smallskip}
Model  &  	   &                &   Parameters &          &                        &           \\
 \hline
\sc{cutoffpl}     &  $N_{\rm H}$$^{\mathrm{a}}$    &     $\Gamma$          &   $E_{\rm cut}$$^{\mathrm{b}}$  &   & &  &  Flux$^{\mathrm{c}}$  & $\chi^{2}_{\nu}$/d.o.f. \\
                  &  $6.7_{-0.7}^{+0.7}$ 	&  $0.3_{-0.2}^{+0.2}$ &     $9.1_{-1.7}^{+2.4}$         &  & &   &    $4.8 _{-0.1}^{+0.1}$            & $1.0/263$\\

\sc{bmc*highecut}       &  $N_{\rm H}$$^{\mathrm{a}}$ &    $kT_{\rm BB}$$^{\mathrm{b, d}}$     &   $\alpha$$^{\mathrm{d}}$            & $\log(A)$$^{\mathrm{d}}$ &   E$_{\rm cut}$$^{\mathrm{b}}$    & $E_{\rm fold}$$^{\mathrm{b}}$ & Flux$^{\mathrm{c}}$& $\chi^{2}_{\nu}$/d.o.f. \\
                 &  $5.3_{-0.3}^{+0.3}$ &  $1.7_{-0.3}^{+0.3}$ & $1.2_{-1.2}^{+0.2}$  & $5.7_{-13.7}^{+2.3}$  &  $23_{-8}^{+5}$ & $20_{-14}^{+10}$ & $4.2_{-0.1}^{+0.1}$   & $1.0/260$\\

\sc{Comptt}$^{\mathrm{e}}$ &  $N_{\rm H}$$^{\mathrm{a}}$ &  $kT_0$$^{\mathrm{b}}$  & $kT_{\rm e}$$^{\mathrm{b}}$ & $\tau$  & & & Flux$^{\mathrm{c}}$& $\chi^{2}_{\nu}$/d.o.f. \\
            &  $4.5_{-0.5}^{+0.6}$       &  $1.4_{-1.0}^{+0.1}$ & $6.2_{-1.1}^{+0.6} $   & $10_{-1}^{+5}$     &  & &  $4.6_{-1.8}^{+0.5}$           & $1.0/262 $\\
 \noalign{\smallskip}
  \hline
  \end{tabular}
  \end{center}
  \begin{list}{}{} 
 \item[$^{\mathrm{a}}$]{Absorbing column density is in units of $10^{22}$ cm$^{-2}$.}
   \item[$^{\mathrm{b}}$]{High energy cutoff ($E_{\rm cut}$), electron temperature ($kT_{\rm e}$), 
seed photons temperature ($kT_0$) and  the blackbody color temperature $kT_{\rm BB}$ are all in units of keV.}
  \item[$^{\mathrm{c}}$]{Unabsorbed 1--100\,keV flux is in units of $10^{-9}$ erg cm$^{-2}$ s$^{-1}$.}
 \item[$^{\mathrm{d}}$]{$kT_{\rm BB}$ is the blackbody color temperature of the seed photons, $\alpha$ 
is the spectral index and Log(A) is the illumination parameter.} 
 \item[$^{\mathrm{e}}$]{Assuming a spherical geometry.}
  \end{list}
  \end{table*} 


\bsp

\label{lastpage}


\begin{thebibliography}{30}
\expandafter\ifx\csname natexlab\endcsname\relax\def\natexlab#1{#1}\fi

\bibitem[{{Barthelmy} {et~al.}(2008){Barthelmy}, {Krimm}, {Markwardt},
  {Palmer}, \& {Ukwatta}}]{Barthelmy2008GCN7419}
{Barthelmy}, S.~D., {Krimm}, H.~A., {Markwardt}, C.~B., {Palmer}, D.~M., \&
  {Ukwatta}, T.~N. 2008, GRB Coordinates Network, 7419, 1

\bibitem[{{Bird} {et~al.}(2009){Bird}, {Bazzano}, {Hill}, {McBride}, {Sguera},
  {Shaw}, \& {Watkins}}]{Bird2009}
{Bird}, A.~J., {Bazzano}, A., {Hill}, A.~B., {et~al.} 2009, \mnras, 393, L11

\bibitem[{{Bozzo} {et~al.}(2008){Bozzo}, {Campana}, {Stella}, {Falanga},
  {Israel}, {Rampy}, {Smith}, \& {Negueruela}}]{Bozzo2008:atel1493}
{Bozzo}, E., {Campana}, S., {Stella}, L., {et~al.} 2008, \ATel, 1493, 1

\bibitem[{{Buccheri} {et~al.}(1983){Buccheri}, {Bennett}, {Bignami}, {Bloemen},
  {Boriakoff}, {Caraveo}, {Hermsen}, {Kanbach}, {Manchester}, {Masnou},
  {Mayer-Hasselwander}, {Ozel}, {Paul}, {Sacco}, {Scarsi}, \&
  {Strong}}]{Buccheri1983}
{Buccheri}, R., {Bennett}, K., {Bignami}, G.~F., {et~al.} 1983, \aap, 128, 245

\bibitem[{{Burrows} {et~al.}(2005){Burrows}, {Hill}, \& {Nousek et
  al.}}]{Burrows2005:XRTmn}
{Burrows}, D.~N., {Hill}, J.~E., \& {Nousek}, J.~A., {et~al.} 2005, Space Science
  Reviews, 120, 165

\bibitem[{{Grebenev} \& {Sunyaev}(2005)}]{Grebenev2005}
{Grebenev}, S.~A. \& {Sunyaev}, R.~A. 2005, Astronomy Letters, 31, 672

\bibitem[{{Grebenev} \& {Sunyaev}(2008)}]{Grebenev2008:atel1482}
{Grebenev}, S.~A. \& {Sunyaev}, R.~A. 2008, \ATel, 1482, 1

\bibitem[{{in't Zand} {et~al.}(1998){in 't Zand}, {Heise}, {Smith}, {Muller},
  {Ubertini}, \& {Bazzano}}]{zand1998}
{in 't Zand}, J., {Heise}, J., {Smith}, M., {et~al.} 1998, \iaucirc, 6840, 2

\bibitem[{{in't Zand} {et~al.}(2006){in't Zand}, {Jonker}, {Mendez}, \&
  {Markwardt}}]{zand2006:aTel915}
{in't Zand}, J., {Jonker}, P., {Mendez}, M., \& {Markwardt}, C. 2006, \ATel,
  915, 1

\bibitem[{{Jain} {et~al.}(2009){Jain}, {Paul}, \& {Dutta}}]{Jain2009:16479}
{Jain}, C., {Paul}, B., \& {Dutta}, A. 2009, \mnras, in press, arXiv:0903.5403

\bibitem[{{Negueruela} \& {Schurch}(2007)}]{Negueruela2007sax1818}
{Negueruela}, I. \& {Schurch}, M.~P.~E. 2007, \aap, 461, 631

\bibitem[{{Negueruela} \& {Smith}(2006)}]{Negueruela2006:aTel831}
{Negueruela}, I. \& {Smith}, D.~M. 2006, \ATel, 831, 1

\bibitem[{{Negueruela} {et~al.}(2006){Negueruela}, {Smith}, {Harrison}, \&
  {Torrej{\'o}n}}]{Negueruela2006}
{Negueruela}, I., {Smith}, D.~M., {Harrison}, T.~E., \& {Torrej{\'o}n}, J.~M.
  2006, \apj, 638, 982

\bibitem[{{Paizis} {et~al.}(2006){Paizis}, {Farinelli}, {Titarchuk},
  {Courvoisier}, {Bazzano}, {Beckmann}, {Frontera}, {Goldoni}, {Kuulkers},
  {Mereghetti}, {Rodriguez}, \& {Vilhu}}]{Paizis2006}
{Paizis}, A., {et~al.} 2006, \aap, 459, 187


\bibitem[{{Romano} {et~al.}(2006){Romano}, {Campana}, {Chincarini}}]{Romano2006:060124mn}
{Romano}, P., {Campana}, S., {Chincarini}, G., {et~al.} 2006, \aap, 456, 917


\bibitem[{{Romano} {et~al.}(2009{\natexlab{a}}){Romano}, {Sidoli}, {Cusumano},
  {Evans}, {Ducci}, {Krimm}, {Vercellone}, {Page}, {Beardmore}, {Burrows},
  {Kennea}, {Gehrels}, {La Parola}, \& {Mangano}}]{Romano2009:sfxts_paper08408}
{Romano}, P., {Sidoli}, L., {Cusumano}, G., {et~al.} 2009{\natexlab{a}},
  \mnras, 392, 45

\bibitem[{{Romano} {et~al.}(2009{\natexlab{b}}){Romano}, {Sidoli}, {Cusumano},
  {La Parola}, {Vercellone}, {Pagani}, {Ducci}, {Mangano}, {Cummings}, {Krimm},
  {Guidorzi}, {Kennea}, {Hoversten}, {Burrows}, \&
  {Gehrels}}]{Romano2009:sfxts_paperV}
{Romano}, P., {Sidoli}, L., {Cusumano}, G., {et~al.} 2009{\natexlab{b}},
  \mnras, in press, arXiv:0907.1289 

\bibitem[{{Romano} {et~al.}(2009{\natexlab{c}}){Romano}, {Sidoli}, {Cusumano},
  {Vercellone}, {Mangano}, \& {Krimm}}]{Romano2009:11215_2008}
{Romano}, P., {Sidoli}, L., {Cusumano}, G., {et~al.} 2009{\natexlab{c}}, \apj,
  696, 2068

\bibitem[{{Romano} {et~al.}(2009{\natexlab{d}}){Romano}, {Sidoli}, {Krimm},
  {Chester}, {Evans}, {Barthelmy}, {Vercellone}, {La Parola}, {Mangano},
  {Kennea}, {Burrows}, \& {Gehrels}}]{Romano2009:atel2044}
{Romano}, P., {Sidoli}, L., {Krimm}, H.~A., {et~al.} 2009{\natexlab{d}}, \ATel,
  2044, 1

\bibitem[{{Romano} {et~al.}(2008){Romano}, {Sidoli}, {Mangano}, {Vercellone},
  {Kennea}, {Cusumano}, {Krimm}, {Burrows}, \&
  {Gehrels}}]{Romano2008:sfxts_paperII}
{Romano}, P., {Sidoli}, L., {Mangano}, V., {et~al.} 2008, \apjl, 680, L137

\bibitem[{{Romano} et~al.}(2007){Romano}, {Sidoli}, {Mangano}, {Mereghetti} \& {Cusumano}]{Romano2007}
{Romano} P.,  {Sidoli} L.,  {Mangano} V.,  {Mereghetti} S.,    {Cusumano} G., 2007, \aap, 469, L5

\bibitem[{{Scargle}(1982)}]{Scargle1982}
{Scargle}, J.~D. 1982, \apj, 263, 835

\bibitem[{{Sguera} {et~al.}(2005){Sguera}, {Barlow}, {Bird}, {Clark}, {Dean},
  {Hill}, {Moran}, {Shaw}, {Willis}, {Bazzano}, {Ubertini}, \&
  {Malizia}}]{Sguera2005}
{Sguera}, V., {Barlow}, E.~J., {Bird}, A.~J., {et~al.} 2005, \aap, 444, 221

\bibitem[{{Sguera} {et~al.}(2007){Sguera}, {Hill}, {Bird}, {Dean}, {Bazzano},
  {Ubertini}, {Masetti}, {Landi}, {Malizia}, {Clark}, \& {Molina}}]{Sguera2007}
{Sguera}, V., {Hill}, A.~B., {Bird}, A.~J., {et~al.} 2007, \aap, 467, 249

\bibitem[{{Sidoli}(2009)}]{Sidoli2009:cospar}
{Sidoli}, L. 2009, Advances in Space Research, 43, 1464

\bibitem[{{Sidoli} {et~al.}(2006){Sidoli}, {Paizis}, \&
  {Mereghetti}}]{SidoliPM2006}
{Sidoli}, L., {Paizis}, A., \& {Mereghetti}, S. 2006, \aap, 450, L9

\bibitem[{{Sidoli} {et~al.}(2009{\natexlab{a}}){Sidoli}, {Romano}, {Ducci},
  {Cusumano}, {Mangano}, {Paizis}, {Krimm}, {Vercellone}, {Burrows}, {Kennea},
  \& {Gehrels}}]{Sidoli2009:sfxts_paperIV}
{Sidoli}, L., {Romano}, P., {Ducci}, L., {et~al.} 2009{\natexlab{a}}, \mnras,
  in press, arXiv:0905.2815

\bibitem[{{Sidoli} {et~al.}(2009{\natexlab{b}}){Sidoli}, {Romano}, {Mangano},
  {Cusumano}, {Vercellone}, {Kennea}, {Paizis}, {Krimm}, {Burrows}, \&
  {Gehrels}}]{Sidoli2009:sfxts_paperIII}
{Sidoli}, L., {Romano}, P., {Mangano}, V., {et~al.} 2009{\natexlab{b}}, \apj,
  690, 120

\bibitem[{{Sidoli} {et~al.}(2008){Sidoli}, {Romano}, {Mangano}, {Pellizzoni},
  {Kennea}, {Cusumano}, {Vercellone}, {Paizis}, {Burrows}, \&
  {Gehrels}}]{Sidoli2008:sfxts_paperI}
{Sidoli}, L., {Romano}, P., {Mangano}, V., {et~al.} 2008, \apj, 687, 1230

\bibitem[{{Titarchuk}(1994)}]{Titarchuk1994}
{Titarchuk}, L. 1994, \apj, 434, 570

\bibitem[{{Titarchuk} {et~al.}(1996){Titarchuk}, {Mastichiadis}, \&
  {Kylafis}}]{TMK1996}
{Titarchuk}, L., {Mastichiadis}, A., \& {Kylafis}, N.~D. 1996, \aaps, 120, C171

\bibitem[{{Vaughan} {et~al.}(1994){Vaughan}, {van der Klis}, {Wood}, {Norris},
  {Hertz}, {Michelson}, {van Paradijs}, {Lewin}, {Mitsuda}, \&
  {Penninx}}]{Vaughan1994}
{Vaughan}, B.~A., {van der Klis}, M., {Wood}, K.~S., {et~al.} 1994, \apj, 435,
  362


\bibitem[\protect\citeauthoryear{{Vaughan}, {Goad}, {Beardmore}, {O'Brien},
  {Osborne}, {Page}, {Barthelmy} \& {Burrows}}{{Vaughan}
  et~al.}{2006}]{vaughan2006:050315mn}
{Vaughan} S.,  {Goad} M.~R.,  {Beardmore} A.~P., et al.,  2006, \apj,
  638, 920


\bibitem[{{Zurita Heras} \& {Chaty}(2009)}]{Zurita2009}
{Zurita Heras}, J.~A. \& {Chaty}, S. 2009, \aap, 493, L1

\end{thebibliography}
\end{document}